\documentclass[prl,preprint,groupedaddress,showpacs]{revtex4-1}

\usepackage{amsmath}
\usepackage{amssymb}
\usepackage{amsfonts}
\usepackage{mathrsfs}
\usepackage{epsfig}
\usepackage{bm}

\begin{document}
 
\title{Magnetic control of particle-injection in plasma based accelerators}


\author{J. Vieira$^1$}
\author{S. F. Martins$^1$}
\author{V. B. Pathak$^1$}
\author{R. A. Fonseca$^{1,2}$}
\author{W.B. Mori$^3$}
\author{L. O. Silva$^1$}

\affiliation{$^1$GoLP/Instituto de Plasmas e Fus\~{a}o Nuclear-Laborat\'orio Associado,  Instituto Superior T\'{e}cnico, 1049-001 Lisboa, Portugal}
\affiliation{$^2$DCTI/ISCTE Lisbon University Institute, 1649-026 Lisbon, Portugal}
\affiliation{$^3$Department of Physics and Astronomy, UCLA, Los Angeles, California 90095, USA}


\today

\begin{abstract}
The use of an external transverse magnetic field to trigger and to control electron self-injection in laser- and particle-beam driven wakefield accelerators is examined analytically and through full-scale particle-in-cell simulations. A magnetic field can relax the injection threshold and can be used to control main output beam features such as charge, energy, and transverse dynamics in the ion channel associated with the plasma blowout. It is shown that this mechanism could be studied using state-of-the-art magnetic fields in next generation plasma accelerator experiments.
\end{abstract}

\maketitle


Plasma based acceleration (PBA) has the potential to lead to a future generation of compact particle accelerators~\cite{bib:tajima_prl_1979,bib:lu_prstab_2007,bib:martins_natphys_2010} for high energy physics applications, and also to compact light sources for medical applications~\cite{bib:patel_nature_2007} and material science. In state-of-the-art laser wakefield acceleration (LWFA) experiments, relativistic plasma waves capable to self-inject and accelerate 1~GeV class electrons~\cite{bib:leemans_natphys_2006} are excited in the so-called bubble or blowout regime~\cite{bib:pukhov_apb_2002,bib:lu_prl_2006}. In contrast, the self-injection conditions are not easily met in current plasma wakefield acceleration (PWFA) experiments, since the bubble radius is not as large. Nevertheless, in recent PWFA experiments, some electrons had their energy doubled from 42 GeV to 85 GeV in 85~cm~\cite{bib:blumenfeld_nat_2007}.


The output controllability is a major challenge for the use of PBA in several applications. To this end, new concepts have emerged such as counter- and cross-propagating laser pulses for LWFA~\cite{bib:faure_nature_2006}, short plasma down-ramps~\cite{bib:geddes_prl_2008}, ionization induced trapping~\cite{bib:pak_prl_2010}, and evolving bubbles~\cite{bib:kalmykov_prl_2009}. With these methods, the charge and energy of self-injected bunches can be adjusted.

In this Letter we propose a novel scheme that uses static transverse magnetic fields to trigger and to control the self-injection in the LWFA or PWFA. The trapping occurs off-axis leading to synchronized betatron oscillations, which could improve the quality of x-ray emission by the magnetically injected electrons~\cite{bib:whittum_prl_1990}. The output energies can also be controlled by adjusting the longitudinal injection position. The scaling law for the magnetic field induced injection is determined with the appropriate Hamiltonian, and illustrated with particle-in-cell (PIC) simulations in OSIRIS~\cite{bib:fonseca_book}. For the next generation PBAs aiming at producing multi-10~GeV electron bunches in controlled injection scenarios~\cite{bib:lu_prstab_2007,bib:martins_natphys_2010}, our scheme requires external B-fields as low as $5~\mathrm{T}$.     

The use of magnetic fields in PBAs was first explored in the so-called \emph{surfatron}~\cite{bib:katsouleas_prl_1983}. The role of magnetic fields in PBAs was further examined in~\cite{bib:ren_pop_2004}, while Ref.~\cite{bib:hosokai_prl_2006} showed that external longitudinal magnetic fields increase the trapped charge. Our scheme uses an external magnetic field perpendicular to the driver velocity to generate controlled off-axis injection bursts.



Self-injection can be investigated through the Hamiltonean $H = \sqrt{m_e^2 c^4+(\mathbf{P}+e \mathbf{A}/c)^2}-e \phi$ for background electrons, where $\mathbf{P}=\mathbf{p}-e \mathbf{A}$ and $\mathbf{p}$ are the canonical and linear momenta, $-e$ the charge of the electron, and $\mathbf{A}$ and $\phi$ the plasma vector and scalar potentials. Unless stated, normalized units are adopted henceforth. Durations are normalized to the inverse of the plasma frequency $\omega_p=\sqrt{4 \pi n_0 e^2/m_e}$, lengths to $c/\omega_p$, velocities to $c$, momenta to $m_e c$, and charge to $e$, with $n_0$ the background plasma density, and $m_e$ the electron mass. Vector and scalar potentials are normalized to $e/m_e c^2$ and to $e/m_e c$. The normalized B-field is given by $\omega_c/\omega_p = e |\mathbf{B}|/m_e \omega_p$, and $\omega_c$ is the cyclotron frequency.

In the co-moving frame $(x=x,y=y,\xi=v_{\phi}t-z,s=z)$, the Hamiltonian is $\mathcal{H}=H-v_{\phi} P_{\|}$, where $v_{\phi}$ is the plasma wave phase velocity, and $P_{\|}$ is the longitudinal canonical momentum. In this frame $\mathrm{d}\mathcal{H}/\mathrm{d} \xi = \left(1-v_{z}/v_{\phi}\right)^{-1}\left(\mathbf{v}\cdot \partial \mathbf{A}/\partial s-\partial \phi/\partial s\right)$~\cite{bib:luthesis, bib:pak_prl_2010}, where $v_z$ is the electron velocity parallel to the driver velocity. Thus, defining the wake potential as $\psi=\phi-v_{\phi} A_z$~\cite{bib:luthesis}, and since $v_z=v_{\phi}$ for trapped electrons, $1+\Delta \psi = \gamma/\gamma_{\phi}^2 - \Delta{\mathcal{H}}$ at the instant of injection, where $\gamma_{\phi}=(1-v_{\phi}^2)^{-1/2}$ is the relativistic factor of the bubble phase velocity, and $\Delta \psi=\psi_f-\psi_i$ ($\Delta{\mathcal{H}}=\mathcal{H}_f-\mathcal{H}_i$) is the difference between the $\psi$ ($\mathcal{H}$) of the electron at its trapping and initial position. This trapping condition is consistent with the conditions from other  electron self-injection mechanisms~\cite{bib:luthesis, bib:kostyukov_prl_2009,bib:kalmykov_prl_2009,bib:pak_prl_2010}, and is also valid when external fields are present. Considering that $\psi^{\mathrm{ext}}$ is related to the external field, and denoting the plasma wave contribution to $\psi$ by $\psi^{\mathrm{pl}}$, the trapping condition becomes $1+\Delta \psi^{\mathrm{pl}} = \gamma/\gamma_{\phi}^2-\Delta{\mathcal{H}}-\Delta \psi^{\mathrm{ext}}$.

The trapping condition derived above shows that self-injection can be relaxed by the presence of external fields.  In the absence of external fields ($\Delta \psi^{\mathrm{ext}}=0$) and in the quasi-static approximation~\cite{bib:kostyukov_prl_2009}, such that $\Delta \mathcal{H}=0$, self-injection occurs when $\Delta \psi^{\mathrm{pl}}$ approaches $\Delta \psi^{\mathrm{pl}} = -1$~\cite{bib:tsung_prl_2004,bib:pak_prl_2010,bib:kostyukov_prl_2009}. This condition can be relaxed (i.e. lower $|\Delta \psi|$ can still lead to injection) during the expansion of the wakefield~\cite{bib:kalmykov_prl_2009} which contributes with finite $\Delta{\mathcal{H}}$. In the presence of external fields, self-injection may be controlled by acting on $\Delta{\mathcal{H}}$ and $\Delta \psi^{\mathrm{ext}}$. To illustrate the latter mechanism, an external constant magnetic field $B_y$ pointing in the positive y-direction is considered, and described as $\psi^{\mathrm{ext}}=-A^{\mathrm{ext}}_z=B_y x$. We also assume that the B-field rises from zero to $B_{y0}$ in a length $L^{\mathrm{ramp}}$, is constant for $L^{\mathrm{flat}}$, and vanishes in $L^{\mathrm{ramp}}$.

We start by investigating trapping in the uniform B-field region, for which the trapping condition reduces to $1+\Delta \psi^{\mathrm{pl}}=-B_{y0} \Delta x$ when $\gamma_{\phi}\rightarrow \infty$, and where $\Delta x=x_f-x_i$ is the difference between the final and initial electron position in the x direction. A conservative threshold B field for injection can be retrieved assuming that $\Delta \psi^{\mathrm{pl}}=0$, and that the wake is relatively unperturbed by the B-field (i.e. $\omega_c/\omega_p \ll 1$). In LWFA matched propagation regimes, the B-injected electrons originate at a distance $x_i\simeq r_b$ from the axis, and they are trapped with $x_f\simeq 0$ ~\cite{bib:tsung_prl_2004}. Thus, for a trapped electron $\Delta x\simeq-r_b \sin{\theta} = -2 \sqrt{a_0} \sin{\theta}$, where $r_b=2 \sqrt{a_0}$ is the blowout radius~\cite{bib:lu_prl_2006}, $\theta$ the angle between the plane of the trajectory of the electron with the B-field, and $a_0$ is the normalized laser vector potential. Then, the threshold B-field for injection becomes:
\begin{equation}
\label{eq:threshold}
B^{\mathrm{t}}=\frac{1}{2 \sin{\theta} \sqrt{a_0}}.
\end{equation}
Eq.~(\ref{eq:threshold}) illustrates that the external B-field leads to localized off-axis injection in a well defined angular region. We note, however, that since typically $\Delta\psi\lesssim -1$, the assumption $\Delta \psi=0$ significantly overestimates the required B-field for injection. In fact as we will show in 3D OSIRIS PIC simulations of the PWFA, the threshold B-field for injection is significantly lower and can be within the reach of current technology~\cite{bib:pollock_rsi_2006}. 

The localized injection in the transverse x-y space can also be interpreted in terms of the Larmor rotation of the plasma electrons. Since backward moving electrons in the z-direction rotate anticlockwise for $B_{y0}>0$, they are bent towards the axis for $x>0$, entering the accelerating and focusing region of the bubble with larger $p_{\|}$, thus facilitating self-injection. For $x<0$, electrons move away from the axis, preventing injection. Furthermore, the electrons moving in the plane $x \simeq 0$ are much less disturbed. Hence, the B-field leads to off-axis injection in a defined angular region. 

If $B_y\gg B^{\mathrm{t}}$ the plasma wave structure can be significantly modified, which may reduce $|\Delta \psi|$ and $|\Delta x|$, and suppress injection. In such cases, however, injection can still occur because the B-field depends on $\xi$ (contribution of $\Delta \mathcal H$ and $\Delta \psi^{\mathrm{ext}}$ to the trapping condition). Specifically, as the B-field begins to decrease in the down ramp region then $\psi^{\mathrm{pl}}$ returns to its unperturbed state, and $r_b$ gets larger, effectively lowering $v_{\phi}$. In the down ramp B-field regions injection also occurs off-axis and in an angular region that is identical to that associated with the uniform B-field regions. We stress that for the scenarios presented here, the majority of the charge is injected in the B-field down ramps. This is the most important mechanism in the present configuration, although for sufficiently large $L^{\mathrm{ramp}}$ the mechanisms associated with constant B-fields may also lead to self-injection.

According to the trapping condition in the presence of external fields, higher injection rates occur for higher plasma wave expansion rates, or equivalently for higher B-field down-ramp gradients, i.e. for shorter ramps or higher B's which provide a larger $|\mathbf{v}\cdot \partial \mathbf{A}/\partial s - \partial \phi/\partial s|$ and hence larger $|\Delta \mathcal{H}|$. Physically, both tend to lower the phase velocity due to the accordion effect~\cite{bib:kat_pra_1986}. We note that $L^{\mathrm{ramp}}$ can change by modifying the spatial extent of the B-field, and by varying $n_0$.

\begin{figure}[h!tbp]
\begin{center}
\includegraphics[width=0.6 \columnwidth]{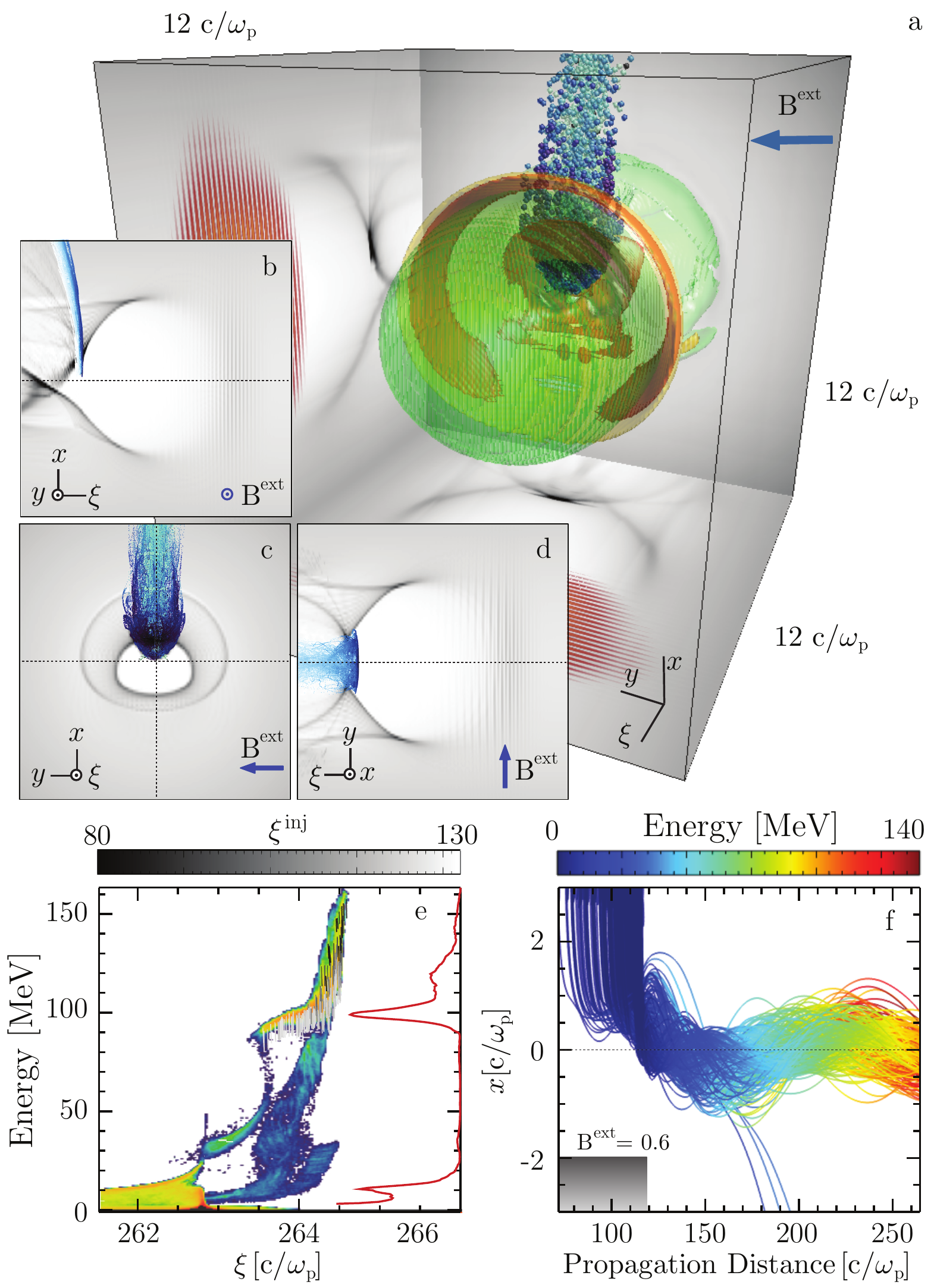}
\caption{\label{fig:bubble3d} 3D OSIRIS simulation of a magnetized LWFA. \textbf{a}. Electron density iso-surfaces in the uniform B field region. \textbf{b}-\textbf{d}. Density profile associated with the $y=0$, $\xi=0$, and $x=0$ planes of \textbf{a} revealing the off-axis injection. Self-injected electrons (darker dots) are closer to the bubble axis, while the electrons farther from the bubble axis escape from the trapping region (lighter dots). \textbf{e}. Phase-space of the plasma electrons and spectrum (solid line), showing a quasi-monoenergetic ($\sim 6\%$~FWHM spread) electron bunch. \textbf{f}. Trajectories of the self-injected beam particles, with the propagation axis shown by the dashed line. The B-field profile is also represented.}  
\end{center}
\end{figure}

In order to illustrate this controlled injection mechanism we have performed a set of 3D PIC simulations in OSIRIS~\cite{bib:fonseca_book}. We first consider the LWFA scenario, illustrated in Fig.~\ref{fig:bubble3d}. The simulation box with dimensions of $24\times24\times 12~(c/\omega_p)^3$ moves at the speed of light, and is divided into $480\times480\times1200$ cells with $2\times1\times 1$ particles per cell. The ions form an immobile neutralizing background fluid. A linearly polarized laser pulse with central frequency $\omega_0/\omega_p=20$ was used, with a peak vector potential of $a_0=3$, a duration $\omega_p\tau_{\mathrm{FWHM}}=2\sqrt{a_0}$, and a transverse spot size ($W_0$) matched to the pulse duration such that $W_0=c \tau_{\mathrm{FWHM}}$~\cite{bib:lu_prstab_2007}. The plasma density is of the form $n=n_0(z)\left(1+\Delta n r^2\right)$ for $r<10~c/\omega_p$ and $n=0$ for $r>10~c/\omega_p$ with $\Delta n = 4/W_0^4$ being the linear guiding condition, and where $n_0(z)$ is a linear function of $z$ which increases from $n_0=0$ to $n_0=1$ in $50~c/\omega_p$ ensuring a smooth vacuum-plasma transition. At the point where the plasma density reaches its maximum value, a constant external field pointing in the positive y-direction rises with $B_y^{\mathrm{ext}}=\omega_c/\omega_p = 0.6 \sin^2[\pi (z-z_0)/ (2 L^{\mathrm{ramp}})]$, with $L^{\mathrm{ramp}}=10~c/\omega_p$ and $z_0=62~c/\omega_p$, it is constant and equal to $B_{y0}^{\mathrm{ext}}=0.6$ for $L^{\mathrm{flat}}=40~c/\omega_p$ and drops back to zero with $B_y^{\mathrm{ext}}=0.6 \sin^2[\pi (z-z_1)/(2 L^{\mathrm{ramp}})]$ with $z_1=L^{\mathrm{ramp}}+L^{\mathrm{flat}}+z_0$. For a laser with central wavelength of 800 nm, these parameters correspond to a 0.473 J laser pulse, with $W_0=8.68~\mu$m, and $\tau_{\mathrm{FWHM}}=28.9~$fs. Moreover, $n_0 = 4.5\times10^{18}~\mathrm{cm}^{-3}$, and $B_{y0}=407~\mathrm{T}$. While this simulation clearly identified key physical mechanisms of the B-field injection using reasonable computational requirements, additional 2D simulations (not shown) revealed that self-injection assisted by magnetic fields can be achieved in the LWFA within state-of-the-art magnetic field generation technology~\cite{bib:pollock_rsi_2006}.

Fig.~\ref{fig:bubble3d}a shows the electron density, the laser projections, and electrons above 10 MeV (spheres). Each quantity is plotted when the front of the laser has just left the B-field, at $\omega_p t = 120$, when the laser evolution~\cite{bib:vieira_njp_2010} can be neglected, corresponding to the wake lying entirely within the downramp region. Fig.~\ref{fig:bubble3d}b-d shows how the B-field can lead to self-injection. Although the B-field can still decrease $r_b$ in the y-$\xi$, x=0 plane, the wake remains symmetric. This can be seen as follows: when an electron is expelled sideways the B-field provides no extra force. However, as the B-field within the wake bends the trajectory backwards in z, the external B-field (in y) produces a force in the negative x direction. This motion then provides a force in the positive z thereby decreasing $r_b$. On the other hand, the wake is asymmetrically modified in the x-z (y=0) plane. Electrons moving backwards in z feel an downward force from the external B-field. This external force reinforces (reduces) the focusing force for electrons with $x>0~(x<0)$. This leads to the sheath structure seen in Fig.~\ref{fig:bubble3d}a. As predicted from the trapping condition, electrons with $\Delta \psi^{ext} <0$ or equivalently $\Delta x <0$ are more easily trapped, thus guaranteeing off axis injection.

The localized trapping provided by the external B-field is also seen in Fig.~\ref{fig:bubble3d}a-c where all electrons above 10 MeV are shown as spheres (dots) at $\omega_p t = 120$. The electrons all reside with $x>0$ and are localized in y as well.  Electrons which eventually reside outsize the wake (large x) are actually defocused by the external B-field. These energetic electrons now move forward in z such that the external B-field leads to a force in the positive x direction. Electrons outside the wake do not feel the focusing force of the wake and therefore are lost. 

In the B-field down-ramp, $r_b$ increases, effectively decreasing $v_{\phi}$, and leading to an additional and stronger injection. The energy spectrum of the electrons at later propagation distances are shown in Fig.~\ref{fig:bubble3d}e, revealing that a quasi mononergetic beam is formed at $\omega_p t = 270$. The total charge of electrons above 50 MeV is 0.15~nC. The streaks in Fig.~\ref{fig:bubble3d}e correspond to the trajectories in Energy-$\xi$ phase space, and are plotted during the time between 240-270~$\omega_p^{-1}$. The color corresponds to their initial value of $\xi$ ($\xi_{\mathrm{inj}}$), then showing that most electrons originate in the downramp region $112<\xi^{\mathrm{inj}}<122$. Fig.~\ref{fig:bubble3d}f shows that self-injected electrons start with $x>0$ forming a beam that has a centroid executing betatron oscillations. 

\begin{figure}[htbp]
\begin{center}
\includegraphics[width=0.8 \columnwidth]{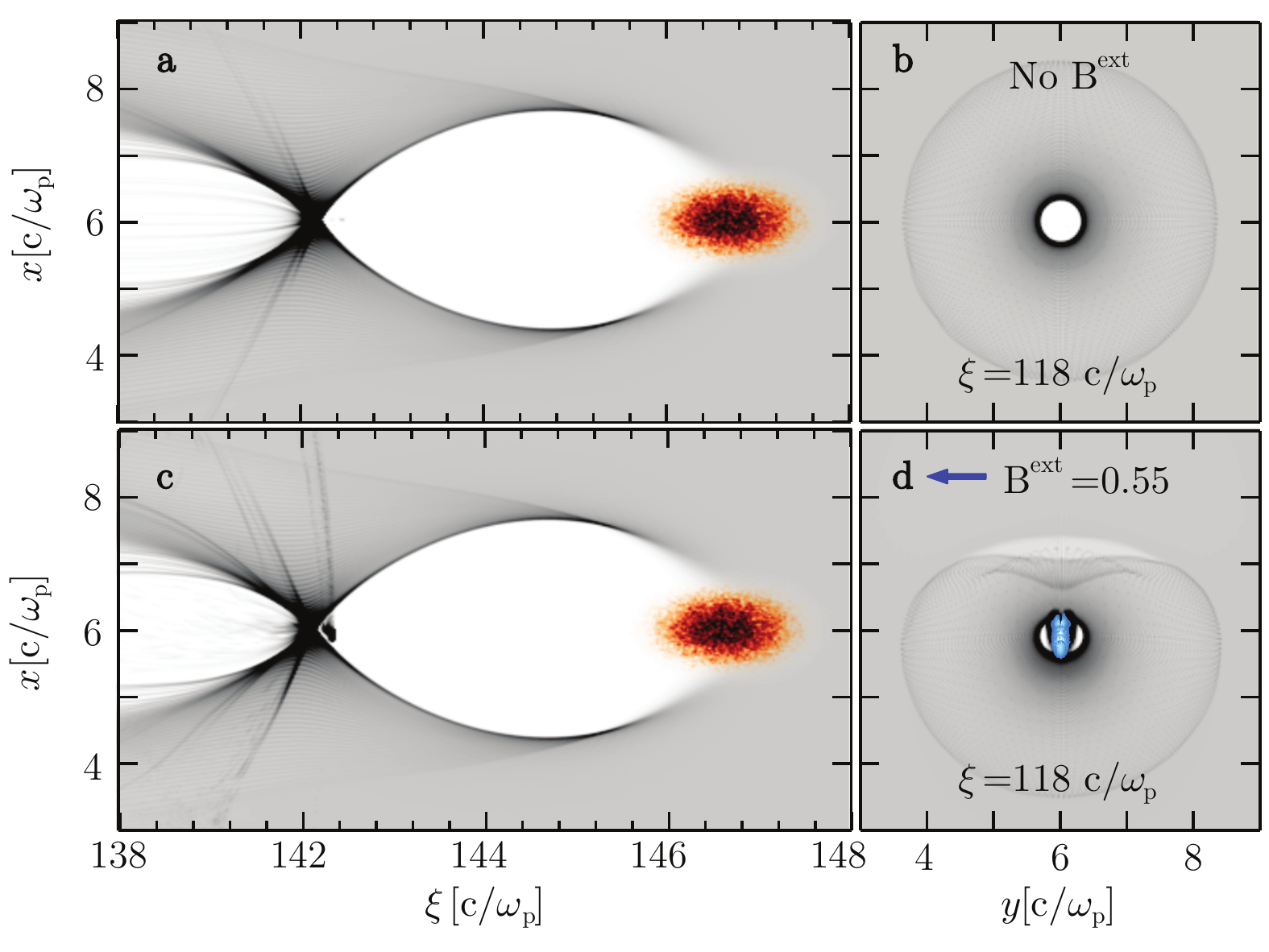}
\caption{\label{fig:pwfa} 3D OSIRIS simulation of B-injection in PWFA (2D slices represented). \textbf{a}. Electron density, and \textbf{b}. transverse density slice at the back of the bubble in the unmagnetized case. \textbf{c}. Electron density after the B-field down-ramp and \textbf{d}. Transverse density slice at the back of the bubble, revealing the trapped particles.}  
\end{center}
\end{figure}

In Fig.~\ref{fig:pwfa} this scheme is applied to PWFA. The energy of the electron beam driver is 30 GeV, and its density profile given by $n_b=n_{b0}\exp\left(-\mathbf{x}_{\perp}^2/(2 \sigma_{\perp}^2)\right)\exp\left(-\xi^2/(2 \sigma_{z}^2)\right)$, where $\sigma_{\perp}=0.3~c/\omega_p$, $\sigma_z=0.5~c/\omega_p$, and $n_{b0}=8.89~n_0$. For $n_0=10^{15}~\mathrm{cm}^{-3}$, this corresponds to $\sigma_{\perp}=50.4~\mu$m, $\sigma_z=84~\mu$m and to a total number of $3\times10^{10}$ electrons, close to the SLAC electron beam. The simulation box is $12\times12\times16~(c/\omega_p)^3$, divided into $480\times480\times640$ cells with $2\times1\times1$ particles per cell for the electron beam and background plasma. The B-field profile is similar to the LWFA case with $B_{y0}^{\mathrm{ext}}=0.55~\omega_c/\omega_p$. For $n_0=10^{15}~\mathrm{cm}^{-3}$, this corresponds to $L^{\mathrm{flat}}=6.8$~mm, $L^{\mathrm{ramp}}=1.7$~mm, and $B_{y0}^{\mathrm{ext}}=5.5~$T, within current technological reach~\cite{bib:pollock_rsi_2006}.

Injection is absent in the unmagnetized PWFA (Fig.~\ref{fig:pwfa}a-b). In the magnetized case, we observed that self-injection occurs only in the B-field downramp, leading to the generation of a 13 pC electron bunch with $0.1~\mathrm{c}/\omega_p=16.8~\mu\mathrm{m}$ long, and $0.2~c/\omega_p=33.6~\mu\mathrm{m}$ wide for $n_0=10^{15}~\mathrm{cm}^{-3}$. As in the LWFA, the deformed structure of the wave only traps plasma electrons located in a narrow angular region (Fig.~\ref{fig:pwfa}c-d), resulting in synchronized betatron trajectories of the B-injected electrons. We note that the external field also deflects the driver by an angle $\delta\simeq B_{y}^{\mathrm{ext}}L/\gamma_{b}$, where $L$ is the total B-field length, and $\gamma_b$ is the electron beam relativistic factor. In our scenario, as $\delta \simeq 6\times10^{\mathrm{-4}}$, the driving beam deflection is negligible, but can still be corrected by placing additional identical external B-fields pointing in the opposite direction. In fact, separating the additional fields by the betatron wavelength can further enhance synchronized injection.

\begin{figure}[htbp]
\begin{center}
\includegraphics[width=\columnwidth]{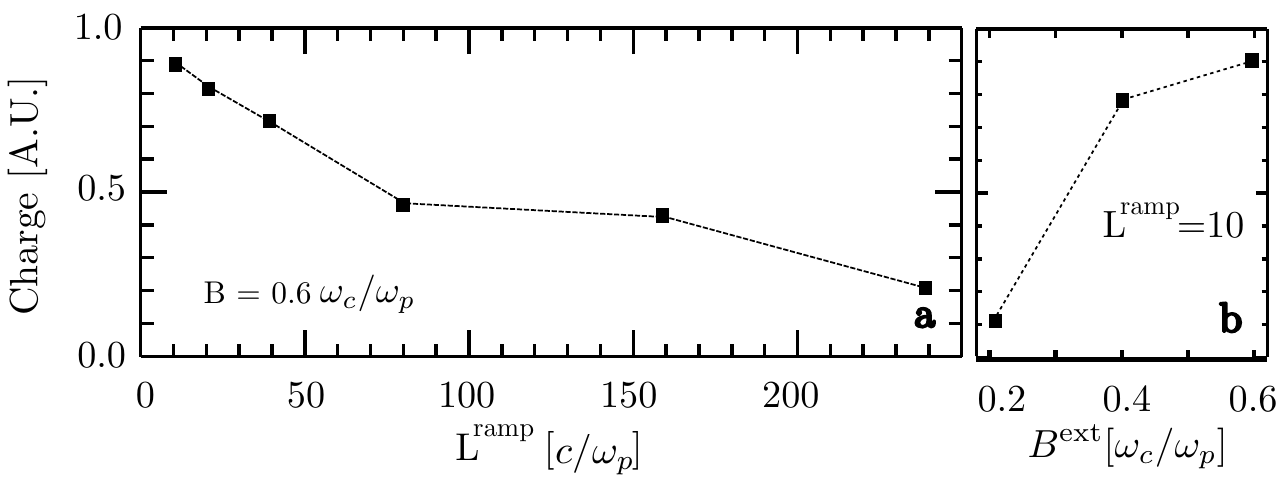}
\caption{\label{fig:results} Parameter scan for the injected charge in the first bucket retrieved from 2D simulations. $\mathbf{a.}$ Down-ramp length scan, revealing that charge decreases with $L^{\mathrm{ramp}}$ $\mathbf{b.}$ Peak B-field scan, revealing the charge increases with $B_y^{\mathrm{ext}}$. }  
\end{center}
\end{figure}

Resorting to 2D simulations, the role of the B-field amplitude, and down ramp length was examined for the LWFA case (Fig.~\ref{fig:results}), starting with the parameters associated with Fig.~\ref{fig:bubble3d}, but using $\omega_0/\omega_p = 50$. Fig.~\ref{fig:results} supports our predictions, showing that the self-injected charge decreases with $L^{\mathrm{ramp}}$ (Fig.~\ref{fig:results}a), and increases with $B_{y0}^{\mathrm{ext}}$ (Fig.~\ref{fig:results}b). The simulations also showed that the self-injected beam radius, energy spread, and beam emittance generally lowers with the applied B-field amplitudes and for shorter ramps. Moreover the beam duration increase for larger B-fields, and for larger field downramps. The injection shutdown occurs for $L^{\mathrm{ramp}}$ larger than a few hundred electron skin-depths, corresponding to the cm-mm scale for plasmas with $n_0=10^{16}-10^{17}~\mathrm{cm}^{-3}$. Finally, these simulations indicate that the threshold B field for injection in the down-ramps is similar to that associated with the uniform B-field region, corresponding to $B_y^{\mathrm{ext}}=1-100~\mathrm{T}$ with $n_0=10^{15}-10^{19}~\mathrm{cm}^{-3}$.

In conclusion, we explored a novel controlled injection mechanism valid for both LWFA and PWFA. This scheme has the potential to generate high-quality beams, since it allows for the tailoring of the injection time period and azimuthal range. It is possible to generate 0.1-1 nC-class electron bunches depending on the plasma and beam density parameters. The plasma, and B-field dependence for the injected charge, and self-injection were estimated analytically and are consistent with the results from PIC simulations in OSIRIS.

Work partially supported by FCT (Portugal) through grants SFRH/BD/22059/2005, PTDC/FIS/111720/2009, and CERN/FP/116388/2010, EC FP7 through LaserLab-Europe/Laptech; UC Lab Fees Research No. 09-LR-05-118764-DOUW, the US DOE under DE-FC02-07ER41500 and DE-FG02-92ER40727, and the NSF under NSF PHY-0904039 and PHY-0936266. Simulations were done on the IST Cluster at IST, on the Jugene supercomputer under a ECFP7 and a DEISA, and on Jaguar computer under INCITE.

\end{document}